\documentclass[12pt,preprint]{aastex}

\shorttitle{The Nature of the Compact X-ray Source in Supernova
Remnant RCW 103} \shortauthors{Li}

\begin{document}

%% LaTeX will automatically break titles if they run longer than
%% one line. However, you may use \\ to force a line break if
%% you desire.

\title{The Nature of the X-ray Source in SNR RCW 103}

\author{Xiang-Dong Li}
\affil{Department of Astronomy, Nanjing University,
    Nanjing 210093, China;\\ lixd@nju.edu.cn}

%% Notice that each of these authors has alternate affiliations, which
%% are identified by the \altaffilmark after each name.  Specify alternate
%% affiliation information with \altaffiltext, with one command per each
%% affiliation.

%% Mark off your abstract in the ``abstract'' environment. In the manuscript
%% style, abstract will output a Received/Accepted line after the
%% title and affiliation information. No date will appear since the author
%% does not have this information. The dates will be filled in by the
%% editorial office after submission.

\begin{abstract}

The discovery of the 6.67 hr periodicity in the X-ray source 1E
161348$-$5055 associated with the supernova remnant RCW 103 has
raised interesting suggestions about the nature of the X-ray source.
Here we argue that in either accreting neutron star or magnetar
model, a supernova fallback disk may be a critical ingredient in
theoretical interpretations of 1E 161348$-$5055. We further
emphasize the effect of fallback disks on the evolution of young
compact objects in various ways, and suggest that even SS 433 could
also be powered by fallback disk accretion process.

\end{abstract}


\section{Introduction}

The enigmatic radio-quiet central compact objects discovered in
supernova remnants (SNRs) have challenged conventional thoughts
that most young neutron stars (NSs) evolve in a manner similar to
Crab-like pulsars \citep[see][for recent reviews]{pav04,kas06}.
The X-ray source 1E 161348$-$5055 (hereinafter 1E1613), the
prototype of the growing radio-quiet objects in SNRs, was first
detected with {\em Einstein} as a faint, unresolved source located
near the center of the SNR RCW 103 \citep{tuo80} with an age of
$\sim 2000$ yr \citep{nug84} and distance of $\sim 3.3$ kpc
\citep{cas75}. From X-ray observations with {\em ASCA},
\citet{got97} found a compact X-ray source inside RCW 103 with the
X-ray luminosity $L_{\rm X}\sim 10^{34}$ ergs$^{-1}$ and the
black-body temperature about 0.6 keV. Further {\em ASCA}
observation of 1E1613, together with archived {\em Einstein} and
{\em ROSAT} data showed that this source manifested an order-of
magnitude decrease in luminosity over four years \citep{got99},
suggesting that this object may be an accreting source rather a
cooling NS. But no X-ray pulsations, radio or optical counter part
were detected. From {\em Chandra} observations and archival {\em
ASCA} data, \citet{gar00} reported the $\sim 6$ hr periodicity of
its flux. The lightcurve taken by {\em XMM-Newton} did not show
the $\sim 6$ hr periodicity but discovered an eclipse about 3 hr
\citep{bec02}. More recent observations with {\em XMM-Newton}
showed a strong, unambiguous periodic modulation at $6.67\pm 0.03$
hr \citep{luc06}.

The origin of this period could be either an orbital or a spin
period. In the former case 1E1613 would be a low-mass X-ray binary
(LMXB) which survived the SN event; in the latter, 1E1613 could be
an isolated magnetar with  magnetic field of order $10^{15}$ G
\citep{luc06}. In each possibility the source remains to be
peculiar and puzzling. In this {\em Letter}, we discuss the nature
of the X-ray source 1E1613 and its implications for young compact
objects, arguing that in both scenarios a fallback disk may be
required to account for the observational characteristics of
1E1613. We further propose that fallback disks may play a critical
role in the evolution of several kinds of young compact stars
including the famous X-ray source SS 433. Throughout this study we
assume that both 1E1613 and RCW 103 originated from the same SN
event.

\section{Accreting NS interpretation}

The long term flux change suggests that 1E1613 may be an accreting
source \citep{got99}. The lightcurve observed by {\em XMM-Newton}
shows for the evidence of an eclipse, indicating 1E 1613 as a NS
in a binary system \citep{bec02}. The possible companion should be
less massive than $0.4\,M_{\sun}$, if it is a normal star, as
indicated by optical/IR observations \citep{pav04,wan07}. Thus
1E1613 could be a young NS accreting from a low-mass companion
star.

The LMXB interpretation meets many difficulties. One of them is the
very low birthrate of LMXBs. It is well known that LMXBs are
extremely rare objects, with a birthrate of $\la 10^{-6}-10^{-7}$
yr$^{-1}$. The birthrate of 1E1613 as an LMXB would be even lower
due to the very low initial mass of its companion - the progenitor
binary is very likely to either merge during the previous common
envelope evolution \citep{jus06}, or be disrupted during the SN
explosion \citep{bra95}. This makes the probability to find an LMXB
like 1E1613 inside a SNR extremely low. Moreover, if 1E1613 is
indeed an LMXB, it is difficult to explain why there is few
association between SNRs and intermediate- or high-mass X-ray
binaries (HMXBs) in the Galaxy, which are more likely to form and to
be observed. Only a single SNR-HMXB association \citep[SS433 and the
SNR W50,][]{gel80} is known in the Galaxy (see discussion below).

Even if the LMXB can survive the common envelope evolution and the
SN explosion, it will take at least $\sim 10^8-10^9$ yr for orbital
decay so that Roche-lobe overflow from the companion starts (the
companion star of mass $\sim 0.4\,M_{\sun}$ will take a time longer
than the age of the Galaxy by nuclear evolution to fill the
Roche-lobe). This timescale is in sharp contrast with the young age
($\sim 2000$ yr) of the SNR RCW 103. \citet{luc06} suggest that the
LMXB may have a significant orbital eccentricity (due to the SN
kick) and hence orbital modulation in the captured mass by the NS
from the companion's wind. We note that the stellar wind mass loss
rate for a $\la 0.4\,M_{\sun}$ dwarf is as low as $\sim
10^{-15}\,M_{\sun}$ yr$^{-1}$, and obviously cannot power the X-ray
luminosity ($\sim 10^{33}-10^{35}$ ergs$^{-1}$) of 1E1613. Thus the
wind from the donor star, if really relevant to the flux modulation,
should be induced by X-ray irradiation from the NS. X-ray
irradiation can affect stars of mass $<1.5\,M_{\sun}$ under certain
circumstances by ionizing the hydrogen at the base of the irradiated
surface layer and disrupting the surface convection zone
\citep{pod91}. According to \citet{pfa03}, the maximum orbital
period for which X-ray irradiation is important is
\begin{equation}
P_{\rm orb, max}\sim 70\,{\rm
days}(\frac{\epsilon\dot{M}_{-8}}{S_{\rm c, 11}})^{3/4},
\end{equation}
where $\dot{M}_{-8}$ is the accretion rate onto the NS in units of
$10^{-8}\,M_{\sun}$ yr$^{-1}$, $S_{\rm c, 11}$ is the critical
X-ray flux for hydrogen ionization in units of $10^{11}$
ergs$^{-1}$cm$^{-2}$, and $\epsilon < 1$ is a factor that takes
into account the geometry of the accretion disk and the star, the
albedo of the star, and the fraction of X-rays that penetrate
below the stellar photosphere \citep{ham93}. Take $L_{\rm X}\sim
10^{34}$ ergs$^{-1}$ for 1E1613, Eq.~(1) yields $P_{\rm orb,
max}<1.7$ hr, still smaller than the 6.67 hr period. However, if
the orbit is highly eccentric, intensive winds might be excited by
X-ray irradiation at periastron. Nevertheless, if future
observations confirm the existence of the binary companion, it
will have important impact on the current theories of NS
formation.

In our opinion, perhaps it is more natural to assume that 1E1613 is
a NS accreting from the fallback disk rather a low-mass companion.
Current stellar evolution models predict that during the core
collapse of massive stars, a considerable amount of the stellar
material will fall back onto the compact, collapsed remnants (NSs or
BHs), usually in the form of an accretion disk
\citep[e.g.][]{woo93}. This point of view is supported by the
discovery of a remnant disk around the anomalous X-ray pulsar 4U
0142$+$61 \citep{wan06}. In this case one does not need to care
about the age contrast in the low-mass donor and young SNR in the
LMXB scenario. The lifetime of the fallback disks is around
$10^4-10^5$ yr. During this time the mass supply may be sufficient
to power the observed X-ray luminosity \citep{cha00}. The flux
outbursts and dips in the light curve could be due to disk
instabilities and occultations by disc structures, respectively
\citep{luc06}. The 6.67 hr periodicity might be related to the
precession of the NS with a hot spot \citep{hel02} or the disk
itself \citep{kat73}.

\citet{pop07} suggested an interesting idea that 1E1613 could be in
a double NS binary system, where the newborn NS which produced the
SNR has a remnant disk around it, and the older NS, identified as
1E1613, accretes periodically from that disk when passing close to
the companion. The situation here is similar to what happens in
Be/X-ray binaries, in which a NS accretes from the disk-like winds
from its Be companion star in an eccentric orbit. In the disk
truncation model \citep{oka01,oka02} the X-ray outbursts in Be/X-ray
binaries are explained as follows. The NS exerts a negative tidal
torque on the viscous decretion disk of the Be star, resulting in
the truncation of the disk. The disk matter would then accumulate in
the outer rings of the disk until the truncation was overcome by the
effects of global one-armed oscillations, disk warping, etc. The
subsequent sudden infall of high-density disk matter onto the NS
causes type II X-ray outbursts. If the tidal truncation is not very
efficient and the disk extends beyond the Roche-lobe of the Be star
at periastron, the matter could be accreted onto the NS during the
periastron passage, resulting in (quasi-)periodic type I bursts.
Calculations by \citet{zha04} have shown that the narrower the
binary system, the more efficient the truncation, since the
truncation efficiency $\propto P_{\rm orb}^{1/3}$. The orbital
period (6.67 hr) of 1E1613, which is much smaller than those of
Be/X-ray binaries, suggests that the flux variations in 1E1613 may
be erratic rather periodic, if caused by the disk truncation effect.

One of the problems for all the accreting NS scenarios is that, to
produce an accretion luminosity in the range observed, the NS
would have a low magnetic field and/or a slow rotation period,
i.e., $P\sim (0.35-2.5)B_{10}^{6/7}$ s, where $B_{10}$ is the NS
magnetic field in units of $10^{10}$ G, so that the accreting
material can overcome the centrifugal barrier \citep{luc06}.
Although these values seem to be peculiar if compared with the
canonical values ($B\sim 10^{12}$ G and $P\sim 10-100$ ms) of
young NSs, there is accumulating evidence for some NSs born
spinning slowly and with a relatively weak magnetic field
\citep{hal07,got07}.

\section{Magnetar interpretation}

\citet{luc06} favor the idea that 1E1613 may be a magnetar
rotating at 6.67 hour with $B> 10^{15}$ G, as the X-ray
variabilities, luminosity and spectral shape would be naturally
explained in the magnetar frame. Even with such field a NS cannot
spin down to 6.7 hours via magneto-dipole radiation (MDR) during
the lifetime of the SNR. Therefore, a SN fallback disk is required
to interact with the NS, providing additional propeller spin-down
torque. Calculations by \citet{luc06} show that a disk of $3\times
10^{-5}\,M_{\sun}$ would have been enough to slow down, over 2000
years, a $B=5\times 10^{15}$ G magnetar to current period,
provided that the initial spin period should be $P_{\rm i}\ga 300$
ms. Magnetic field in magnetars is generally thought to be
generated by turbulent dynamo, whose strength depends on the
star's rotation rate \citep[][see however, Vink \& Kuiper
2006]{dun92}. Such a period seems to be too long for magnetars,
although its possibility cannot be ruled out.

Additionally, the known magnetars include the anomalous X-ray
pulsars (AXPs) and soft gamma-ray repeaters (SGRs), which are
rotating at $\sim 5-10$ s \citep{woo06}. If 1E1613 is a magnetar,
one needs to explain why its period is much longer than those of
AXPs and SGRs.

The mode of interaction between a NS and a surrounding disk is
determined by the location of the inner radius $R_{\rm in}$ of the
disk with respect to the characteristic radii, the corotation
radius $R_{\rm c}=(GMP^2/4\pi^2)^{1/3}$, and the light cylinder
radius $R_{\rm L}=cP/2\pi$, where $M$ is the mass of the NS
\citep[e.g.][]{ill75,lip92}. The position of the inner radius of
the disk can be estimated by comparing the electromagnetic energy
density generated by the NS with the kinetic energy density of the
disk. The NS is expected to be in the propeller and ejector (radio
pulsar) stage if the inner radius of the disk is beyond the
corotation and light cylinder radius, respectively. Since the
kinetic energy density in the disk has the dependence $\propto
r^{-5/2}$ on the radial distance $r$ from the center of the NS,
steeper than the electromagnetic energy density (or radiation
pressure) outside the light cylinder ($\propto r^{-2}$), stable
equilibrium of the disk outside the light cylinder is not allowed,
unless it is beyond the gravitational capture radius
\citep{lip92}.

In their calculations \citet{luc06} have adopted the traditional
estimates of the radiation pressure from a NS by employing a
rotating magnetic dipole in vacuum to generate the electromagnetic
fields. Recently \citet{eks05} derived the electromagnetic energy
density from the global electromagnetic field solution of
\citet{deu55} for obliquely rotating magnetic dipoles. They showed
that the electromagnetic energy density of a rotating dipole makes a
rather broad transition for disk existence across the light cylinder
for small inclination angles. {\em Within this model the disk could
survive beyond the light cylinder even if the radio pulsar activity
turns on}.

To examine to what extent the fallback disks affect the spin
evolution of magnetars, we carried out Monte Carlo simulations of
the evolution of $10^6$ NSs based on the spin-down model presented
in \citet[][see also Li 2002]{lip92}. As the mass of a fallback
disk is not replenished, mass flow rate in the disk declines and
the inner radius of the disk moves out. The NS in this case
generally passes three evolutionary stages. (1) First is the
``ejector" phase, in which the radiative pressure from the NS is
sufficient to keep the surrounding plasma away from the light
cylinder; the NS evolves as a radio pulsar. Here we assume that
once $R_{\rm in}/R_{\rm L}>x_{\rm cr}$, the gaseous disk is
disrupted by the radiation pressure, and the NS is spun down only
by MDR \citep[here $x_{\rm cr}>1$ is a parameter depending on the
magnetic inclination, and is calculated according to][]{eks05}; if
$1<R_{\rm in}/R_{\rm L}<x_{\rm cr}$, the disk is still stable
beyond $R_{\rm L}$, and MDR also works. (2) If $R_{\rm c}<R_{\rm
in}<R_{\rm L}$, the ``propeller" phase follows, in which the
plasma interacts with the neutron star magnetosphere but further
accretion is inhibited by the centrifugal barrier, and the disk
exerts a propeller spin down torque on the NS. (3) For
sufficiently long time of evolution, the NS will enter the
``accretor" phase, in which the inner disk radius becomes smaller
than the corotation radius and steady accretion of the plasma is
allowed.

In our calculations the initial NS magnetic field $B$ is chosen
from a log normal distribution of mean 15 and standard deviation
0.4. We assume that all NSs were born with a surrounding supernova
fallback disk, the initial masses of which $\log(\Delta
M/M_{\sun})$ are distributed uniformly between $-6$ and $-2$. Mass
flow rate through the disk is assumed to decline in a power law
with time, $\dot{M}\propto t^{-1.25}$ \citep{can90,fra02}. We set
the initial spin periods $P_{\rm i}$ and the inclination angle to
be distributed uniformly between 2 ms and 50 ms, and between
$0\degr$ and $90\degr$, respectively. For the propeller torque, we
adopt the expression $T_{\rm prop}=\dot{M}R_{\rm in}^2[\Omega_{\rm
K}(R_{\rm in})-\Omega_{\rm s}]$, where $\Omega_{\rm K}(R_{\rm
in})$ is the Keplerian angular velocity at $R_{\rm in}$ and
$\Omega_{\rm s}$ the angular velocity of the NS \citep[see also
][]{men99}. We stop the calculations at a fiducial time of $2500$
yr, to be compatible with the age of RCW 103.

In Fig.~1 we plot the histogram of the spin periods for the
magnetars. The hatched regions indicate the propeller/accretor
($+45\degr$) and ejector ($-45\degr$) systems, respectively. It is
noted that most ($\sim 99\%$)\footnote{The percentages presented
here are subject to the uncertainties in the initial parameters and
the spin-down mechanisms adopted, and should not be taken very
seriously. } of the magnetars are in the ejector state, having spin
periods of a few seconds. These features are consistent with most of
AXPs and SGRs, which occupy the majority of the magnetar family. The
spin periods of the propeller/accretors, which consist of $\sim
0.6\%$ of the whole population, have a much broader distribution,
peaking around a few of $10^3$ s. 1E1613 may belong to this latter
group. In this case, the puzzling initial spin period $P\ga 300$ ms
does not appear.

\section{Discussion}

The above arguments suggest that in either accreting NS or
magnetar interpretation, a SN fallback disk seems to be favored,
to explain the observational characteristics of 1E1613. Especially
we show that, in the more reasonable magnetar model, it is
possible to model the spin period evolution in 1E1613 with typical
values of the input parameters for a magnetar. This work
emphasizes that fallback disks may considerably influence the
formation and evolution of young compact objects evolved from core
collapse of massive stars. The existence of fallback disks can be
tested by searching for optical/IR emission from the cooler parts
of the disk \citep{per00,wan06}. Recently \citet{wan07} reported
on search for the optical/infrared counterparts to the central
compact objects in four young SNRs including RCW 103, but found
that there is confusion with several faint stars at the position
of 1E1613.

The fallback disk may manifest itself in various ways. Recent X-ray
observations show that some young pulsars, such as the Crab and Vela
pulsars, may have the jet configuration, which suggests the
existence of a disk surrounding the neutron star \citep[][and
references therein]{bla04}. Such disks can influence the braking
indices and timing ages of some young radio pulsars
\citep{men01,mar01,jia05}. A small group of rotating radio
transients (RRATs) were recently reported by \citet{mcl06}. These
objects are characterized by single, dispersed bursts of radio
emission with durations between 2 and 30 ms. \citet{li06} suggested
that these phenomena could be due to the interaction between the NS
magnetosphere and the surrounding SN debris disk. The pattern of
radio pulsar emission depends on whether the disk can penetrate the
light cylinder and efficiently quench the processes of particle
production and acceleration inside the magnetospheric gap; a
precessing disk may naturally account for the switch-on/off behavior
in PSR B1931$+$24 \citep{kra06}. \citet{cor06} also proposed that
the bursty emission by RRATs is due to a disk of circumpulsar
asteroids randomly straying into the magnetosphere.

Fallback disk accretion may also play an important role in producing
the X-ray emission of some ultraluminous X-ray sources (ULXs) that
are possibly associated with SNRs, e.g. IC 342 X-1 \citep{rob03b}
and the ULX in the SNR MF 16 in NGC 6946 \citep{rob03a}. This
association seems to rule out mass transfer in binaries as the main
energy source, since there is little time for the donor stars (if
they exist) to evolve and transfer mass rapidly to the black holes.
A much more plausible explanation is that the black hole is
accreting from the disk originating from the fallback supernova
debris \citep{li03}.

In our Galaxy the X-ray source SS 433 is usually thought to appear
as an ULX if observed face-on. This bizarre object is an X-ray
binary associated with the SNR W50 \citep[see][for
reviews]{mar84,fab04}. It has a 13 d orbital period and is famous
for the presence of its relativistic jets precessing with a
periodicity of about 162 d. There is no consensus about the binary
component masses and the origin of W50. It might be a stellar
windblown bubble produced by the SS 433 jets, as was proposed by
\citet[][see however, Zealy et al. 1980]{beg80}. The companion
star has been suggested to be a supergiant which is filling its
Roche lobe \citep[][but see Barnes et al. 2006]{hil04}, thereby
producing an extremely high-mass transfer that occurs on thermal
time scale with the rate of $\dot{M}\sim 10^{-4}\,M_{\sun}$
yr$^{-1}$ \citep{kin00}. This gives rise to a total luminosity of
$\sim 10^{40}$ ergs$^{-1}$, although the observed unabsorbed
$2-10$ keV X-ray luminosity is $\sim 10^{35}-10^{36}$ ergs$^{-1}$.
However, recent {\em Chandra} observations by \citet{lop06}
suggest the radius of the companion to be about one third of the
Roche lobe radius. A similar conclusion was also reached by
\citet{bri89} with {\em Ginga} observations. If it is correct,
this indicates that the compact object (possibly a black hole) is
likely to be accreting through a SN fallback disk rather Roche
lobe overflow. Stellar wind accretion is also possible. But it is
difficult to account for the high mass accretion rate and the
existence of a stable accretion disk. Additional evidence for
fallback disk accretion in SS 433 my lie in recent {\em
XMM-Newton} observations by \citet{bri05}. These authors find
over-abundances in different elements in the outflowing gas: Si
and S by factors of $\sim 2$, Ni by $\sim 8$. This overabundance
of heavy elements might be a tracer for the past explosion of the
massive progenitor of the compact star, since elemental synthesis
of for example Ni in the inner parts of the accretion disk appears
unlikely due to the extremely high temperatures required. The
fallback disk accretion is the most natural way to convert the SN
ejecta containing the synthesized heavy elements into the outgoing
jets.

In summary, SN fallback disks may account for the observed features
of young compact stars through various ways of interaction between
the disk and the star. Our results motivate further efforts to
detect disks around NSs and BHs in SNRs, and more detailed models of
the disk-star interaction.

\acknowledgments {This work was supported by the Natural Science
Foundation of China under grant numbers 10573010 and 10221001.}

\begin{figure}
 \plotone{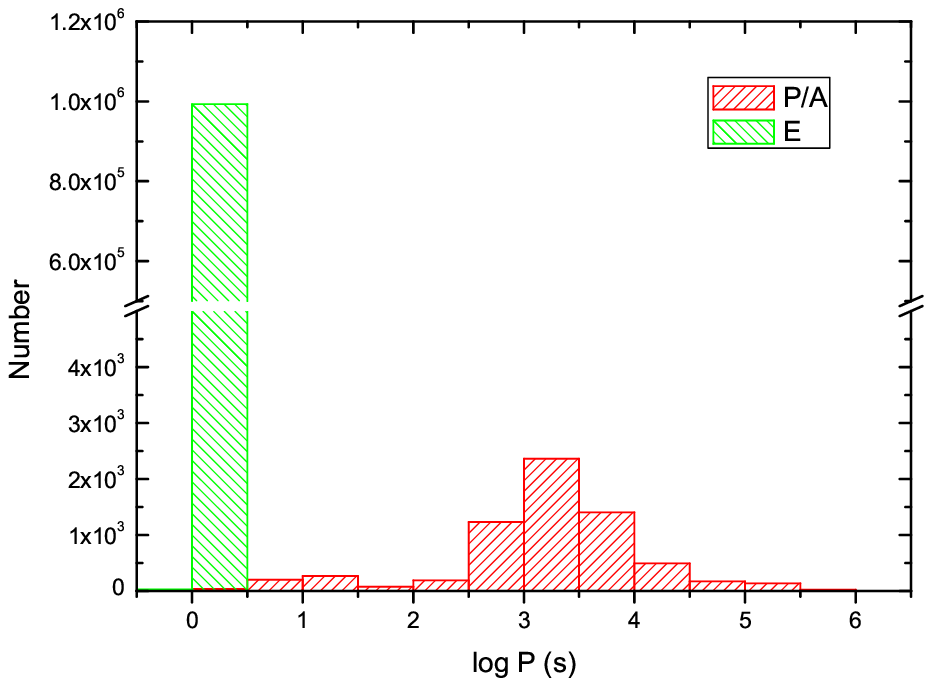}
   \caption{Distribution of the spin periods of magnetars at the age of 2500 yr.
   The hatched regions indicate propeller/accretor ($+45\degr$) and ejector
   ($-45\degr$) systems, respectively.
    }
   \label{}
\end{figure}
\clearpage

\end{document}